\documentclass[12pt,letterpaper]{article}
\usepackage[sort&compress, numbers]{natbib}
\usepackage{graphicx}
\usepackage{amssymb}
\usepackage[margin=1in,top=0.5in,bottom=0.6in]{geometry}
\usepackage{lineno}
\usepackage{enumitem}
\usepackage{aas_macros}
\newcommand{\msun}{M_{\odot}}
\usepackage{hyperref}

\begin{document}

\begin{titlepage}
	\centering
	{\large\bfseries Astro2020 Science White Paper\par}
	\vspace{0.6cm}
	{\scshape\LARGE Supermassive Black-hole Demographics \& Environments With Pulsar Timing Arrays \par}
	\vspace{1cm}
	\includegraphics[width=0.8\textwidth]{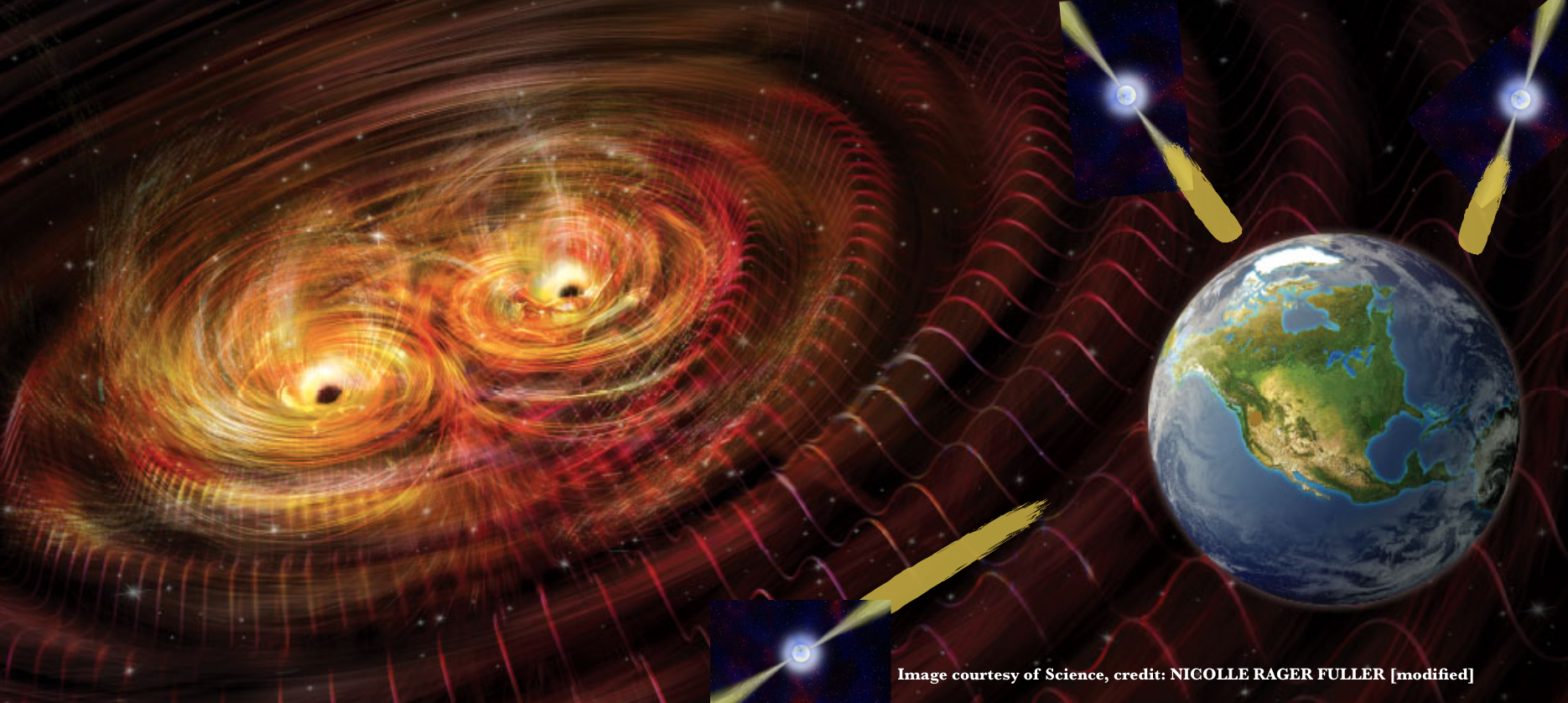}\par\vspace{0.7cm}
	{\large \textbf{Principal authors}} \linebreak
    {\large {\scshape Stephen~R.~Taylor}} {\normalsize\textit{(California Institute of Technology)}}, {\normalsize\href{mailto:srtaylor@caltech.edu}{srtaylor@caltech.edu}}
	\linebreak
	{\large {\scshape\large Sarah~Burke-Spolaor}} {\normalsize\textit{(West Virginia University/Center for Gravitational Waves and Cosmology/CIFAR Azrieli Global Scholar)}}, {\normalsize\href{mailto:sarah.spolaor@mail.wvu.edu}{sarah.spolaor@mail.wvu.edu}} \par
	\vspace{0.6cm}
		{\large {\scshape \textbf{Co-authors}}} 
		\begin{itemize}[noitemsep,topsep=0pt,leftmargin=2cm]
		    \item Paul~T.~Baker \textit{(West Virginia University)}
		    \item Maria Charisi \textit{(California Institute of Technology)} 
		    \item Kristina Islo \textit{(University of Wisconsin-Milwaukee)}
		    \item Luke~Z.~Kelley \textit{(Northwestern University)}
		    \item Dustin~R.~Madison \textit{(West Virginia University)}
		    \item Joseph Simon \textit{(Jet Propulsion Laboratory, California Institute of Technology)}
		    \item Sarah Vigeland \textit{(University of Wisconsin-Milwaukee)}
		\end{itemize}
	\vspace{0.6cm}
	{\normalsize This is one of five core white papers written by members of the NANOGrav Collaboration.} \vspace{-0.5cm}
	\paragraph{\textbf{Related white papers}}
	\begin{itemize}[noitemsep,topsep=0pt]
	    \item \textit{Nanohertz Gravitational Waves, Extreme Astrophysics, And Fundamental Physics With Pulsar Timing Arrays}, J.~Cordes, M.~McLaughlin, et al.
	    \item \textit{Fundamental Physics With Radio Millisecond Pulsars}, E.~Fonseca, et al. 
	    \item \textit{Physics Beyond The Standard Model With Pulsar Timing Arrays}, X.~Siemens, et al.
	    \item \textit{Multi-messenger Astrophysics With Pulsar-timing Arrays}, L.~Z.~Kelley, et al.
	\end{itemize}
	
	\vspace{0.5cm}
	
	\begin{flushleft}
	\noindent \textbf{Thematic Areas:} \hspace*{60pt} 
	$\square$ Planetary Systems \hspace*{10pt} 
	$\square$ Star and Planet Formation \hspace*{10pt}\linebreak
    $\square$ Formation and Evolution of Compact Objects \hspace*{10pt} 
    ${\rlap{$\checkmark$}}\square$ Cosmology and Fundamental Physics \linebreak
    $\square$  Stars and Stellar Evolution \hspace*{10pt} $\square$ Resolved Stellar Populations and their Environments \linebreak 
     ${\rlap{$\checkmark$}}\square$    Galaxy Evolution   \hspace*{45pt} 
     ${\rlap{$\checkmark$}}\square$             Multi-Messenger Astronomy and Astrophysics \hspace*{65pt} \linebreak
    \end{flushleft}
		\vfill
\end{titlepage}

\vspace{-5mm}
\section{Science Opportunity: \textit{Probing the Supermassive \\Black Hole Population With Gravitational Waves}} \label{sec:sec1}
\vspace{-2mm}

With masses in the range $10^6$--$10^9\,\msun$, supermassive black holes (SMBHs) are the most massive compact objects in the Universe. They lurk in massive galaxy centers, accreting inflowing gas, and powering jets that regulate their further accretion as well as galactic star formation. The interconnected growth of SMBHs and galaxies gives rise to scaling relationships between the black-hole's mass and that of the galactic stellar bulge. 

Galaxies grow over cosmic time via gas-accretion from the cosmic web and through merging with other galaxies; the latter leads to the resident SMBHs in each inspiraling to the center of the post-merger remnant, forming a bound pair. The environment surrounding these SMBHs influences their dynamical evolution, affecting how quickly their orbital evolution becomes dominated by gravitational-wave (GW) radiation reaction. Somewhere well within an orbital separation of 1\,pc, the pair will decouple from external influences, and can evolve primarily via the emission of GWs as a pure $2$-body system. Through this decoupling, the GWs frequencies lie in the $\sim 1-100$ nHz band, far below any ground-based (e.g. LIGO, Virgo, KAGRA) or putative space-borne (e.g. LISA) detector. Only pulsar-timing arrays (PTAs) such as NANOGrav (North American Nanohertz Observatory for Gravitational Waves), and potentially precision astrometry missions (e.g. Gaia), can directly probe the decoupling. The ensemble signal from all binaries produces a stochastic background of GWs, whose spectrum encodes their population demographics and dynamical evolution.

In this white paper, we address key questions for our understanding of SMBH populations and galaxy formation, as well as opportunities for multi-messenger nHz-GW astrophysics:
\begin{enumerate}[label=Q\,\arabic*.]
{\bfseries \item How are galaxy properties linked to those of their resident SMBHs?} At moderate redshifts, SMBH masses are inferred through gas or stellar dynamics. These measurements can be prone to biases, measuring all dynamical mass (not just the SMBH) within a certain radius. GW observations \textit{directly} measure the SMBH binary system mass, allowing direct assessment of any scaling relations with a galactic host.
\vspace{-1.75mm}{\bfseries \item How do SMBH pairs evolve from kpc to mpc separations?} Dynamical friction causes SMBHs to sink within a common merger remnant, giving way to repeated stellar-scattering events and circumbinary-disk hardening, then GW orbital decay. We have no conclusive measurements of how long each of these stages lasts, or if there are enough stars to harden the binary to sub-parsec separations. This evolutionary sequence will be addressed directly by nHz-band GW observations, and enhanced at the largest separations by EM observations (see related whitepaper by Kelley et al.).
\vspace{-1.75mm}{\bfseries \item What electromagnetic signatures mark the inspiral and coalescence of SMBHs?}  Paired active nuclei (AGN) can be directly imaged as multiple radio cores, while quasi-variability in nuclear light-curves may indicate binaries in tight orbits. Circumbinary disks may also offer electromagnetic signatures. How will future instruments and surveys work in synergy with nHz GW observations to unveil binary SMBH environments?
\end{enumerate}

\vspace{-8mm}
\section{Science Context}
\vspace{-5mm}
\subsection{Linking Supermassive Black Holes \& Their Host Galaxies}\label{sec:populations}
\vspace{-1mm}

Galaxy growth is hierarchical, occurring through both the accretion of gas and major or minor mergers. Massive galaxies ($M_{\ast} > 10^{11} M_{\odot}$) are overwhelmingly elliptical, with quenched star formation. 
SMBHs in the local Universe exhibit a relationship between their mass and the large-scale observables of their host galaxies \citep{gultekin09} (e.g. velocity dispersion of bulge stars, bulge mass, etc.). The simplest explanation for this is a shared growth history, implying that a central black hole seed is present when the galaxy first forms, then grows along with the galaxy over cosmic time. This growth happens both by the inflow and accretion of gas onto the SMBH, and the coalescence of two of these central SMBHs following a major galaxy merger. The latter forms a binary system along the way to coalescence, emitting GWs at nanohertz frequencies that are undetectable by ground-based (LIGO, Virgo) or planned space-borne (LISA) detectors; only Galactic-scale detectors like PTAs have access to these frequencies.

Factors that influence galaxy evolution (like merger rates and the galaxy stellar mass function) have a knock-on influence on the expected number and brightness of these binary SMBH GW sources. However, the rate of major mergers for massive galaxies across cosmic time is poorly constrained; observations have large discrepancies due to differences in sample selection and merger identification \citep{lotz11}. Even when hydrodynamic cosmological simulations are compared with observed samples, the conclusions for massive galaxies do not converge \citep{rgv+15}. Additionally, there is large scatter in the observed relationship between the central black hole mass and host-galaxy mass, making the underlying physical links between SMBH and galaxy growth difficult to discern.

As mentioned in Sec.~\ref{sec:sec1}, GW observations \textit{directly} probe binary SMBH dynamics, allowing us to robustly answer the above \textbf{Q1}. The stochastic GW background signal measured by PTAs can constrain the relationship between SMBH and host bulge masses \citep{ss16}, and inform the galaxy-galaxy merger rate \citep{chen17} -- see \citep{sbs+18} for an overview of PTA GW astrophysics.

\vspace{-5mm}
\subsection{Dynamical evolution of supermassive black-hole binaries} \label{sec:dynamics}
\vspace{-1mm}
 
    \begin{figure*} 
    \center
    \includegraphics[width=0.75\textwidth]{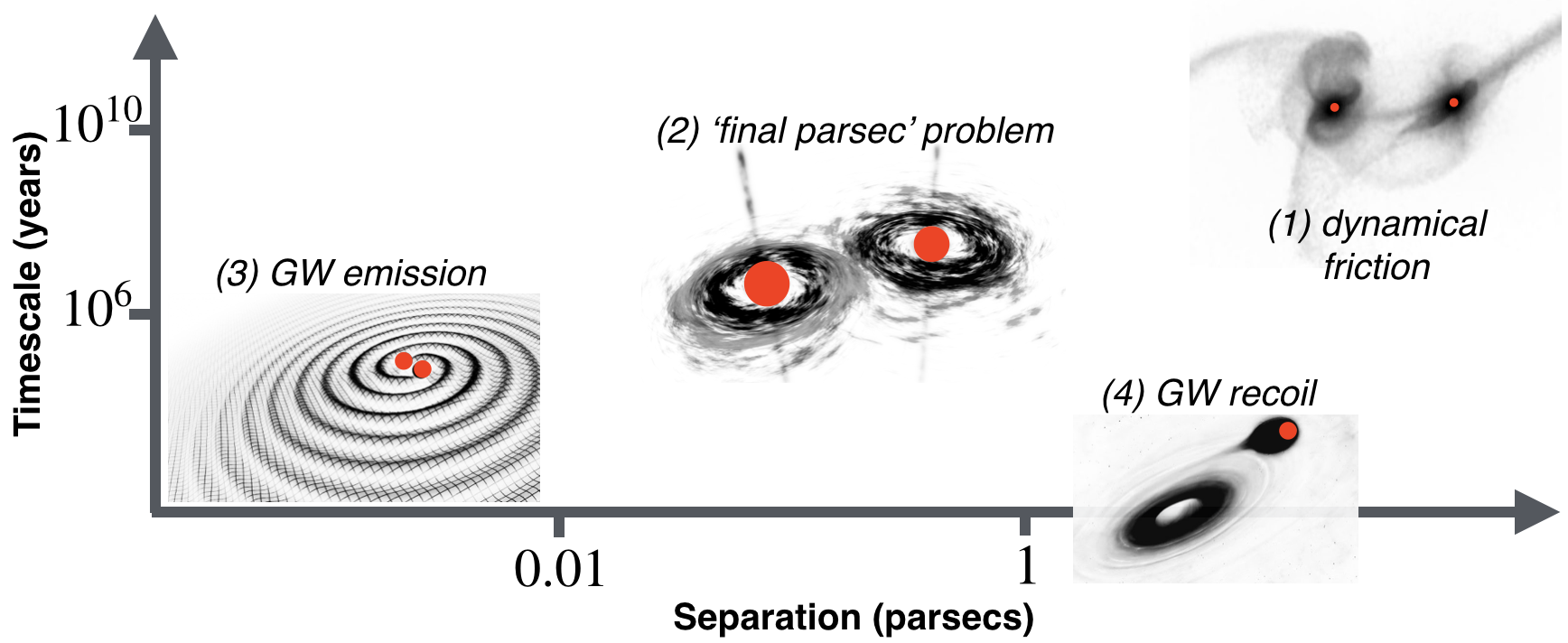}
\vspace{-4mm}
        \caption{Main steps of SMBH evolution following a galaxy merger. Adapted from Ref.~\citep{Komossa_2016,bbr80}.}
        \label{BH_evolution}
\vspace{-3mm}
    \end{figure*}

In Fig.~\ref{BH_evolution}, we show the main steps of evolution for two SMBHs in a post-merger remnant. These are summarized as follows, starting with an uncoupled pair and proceeding inwards:
 
\begin{enumerate}[itemsep=0pt,topsep=0pt,leftmargin=*]
\vspace{-1.5mm}\item \textbf{ Dynamical friction}~($\sim 10^3 - 10\mathrm{pc}$)\\
The SMBHs sink via the drag force induced by relative motion of a massive body through a diffuse medium (e.g., galactic dark matter, stars and gas of the remnant). When adequate energy is extracted from the system, one BH enters the sphere of influence of the other and they form a bound binary system.
\vspace{-1.5mm}
\item \textbf{ Final parsec dynamics}\\
\vspace{-6mm}
\begin{enumerate}[itemsep=0pt,topsep=0pt,leftmargin=*]
\item \textbf{ Stellar loss-cone/3-body scattering}~($\sim 10^{1}$ -- $10^{-1} \, \mathrm{pc}$)\\
The binary orbit continues to decay due to three-body interactions with nearby stars. The population of stars that can interact with the binary may be depleted before the binary reaches the separation at which GW radiation reaction dominates the orbital evolution; this problem is known as the ``final-parsec problem''.
\item \textbf{ Gaseous circumbinary disk interaction}~($\sim 10^{-1}$ -- $10^{-3} \, \mathrm{pc}$)\\
Galaxy mergers deliver large amounts of gas towards the center of the remnant \citep{Barnes1992}.
The gas forms a circumbinary disk, which can catalyze binary evolution while accreting onto the SMBHs and producing bright electromagnetic emission.
\end{enumerate}
\vspace{-1.5mm}
\item \textbf{ Gravitational-wave radiation reaction} ~($\sim 10^{-3}$ pc -- coalescence)\\
At these separations, binary evolution is rapid and driven by the emission of GWs. The final progression to SMBH coalescence encompasses `inspiral' (adiabatic orbital decay), `merger' (complicated GW emission requiring numerical relativity), and `ringdown' (described by BH perturbation theory).
\vspace{-1mm}
\end{enumerate}
The main steps of SMBH evolution have been known for decades \citep{bbr80}, yet large theoretical uncertainties persist. This is unsurprising given the paucity of observed binaries. At sub-kpc separations, a few galaxies have been detected with two active SMBHs (dual AGN). At sub-parsec separations, many candidates have recently emerged from systematic searches for $(1)$ quasar spectra with Doppler-shifted broad emission lines in SDSS \citep{Eracleous2012} and $(2)$ quasars with periodic variability in large time domain surveys \citep{Graham2015,Charisi_2016}. In the coming decade, significant advances are expected on both theoretical and observational fronts; increasingly sophisticated simulations as well as surveys with upcoming telescopes will address the following: 

\begin{itemize}[leftmargin=*]
\vspace{-1.5mm}\item \textbf{Why are galaxies with two SMBHs rare?} Although galaxies merge often, we rarely observe dual and binary SMBHs. This suggests that either SMBH evolution is rapid, and/or they do not typically produce bright electromagnetic signatures \citep{sbs-radiocensus}.
\vspace{-1.5mm}\item \textbf{Do binaries stall at parsec separations?} Recent simulations with improved stellar distributions have shown stellar scattering can effectively evolve a binary. However, these models still cannot model the complete dynamics over the relevant timescales.
\vspace{-1.5mm}\item \textbf{Can gas catalyze the binary evolution in `dry' mergers?} Massive ellipticals have relatively low cold-gas fractions. However, large quantities of hot gas can still be driven towards the galactic nuclei, and may affect the hardening process.
\end{itemize}

\vspace{-8mm}
\subsection{Gravitational-wave Signatures} \label{sec:signatures}
\begin{figure*} 
    \center
    \includegraphics[width=0.95\textwidth]{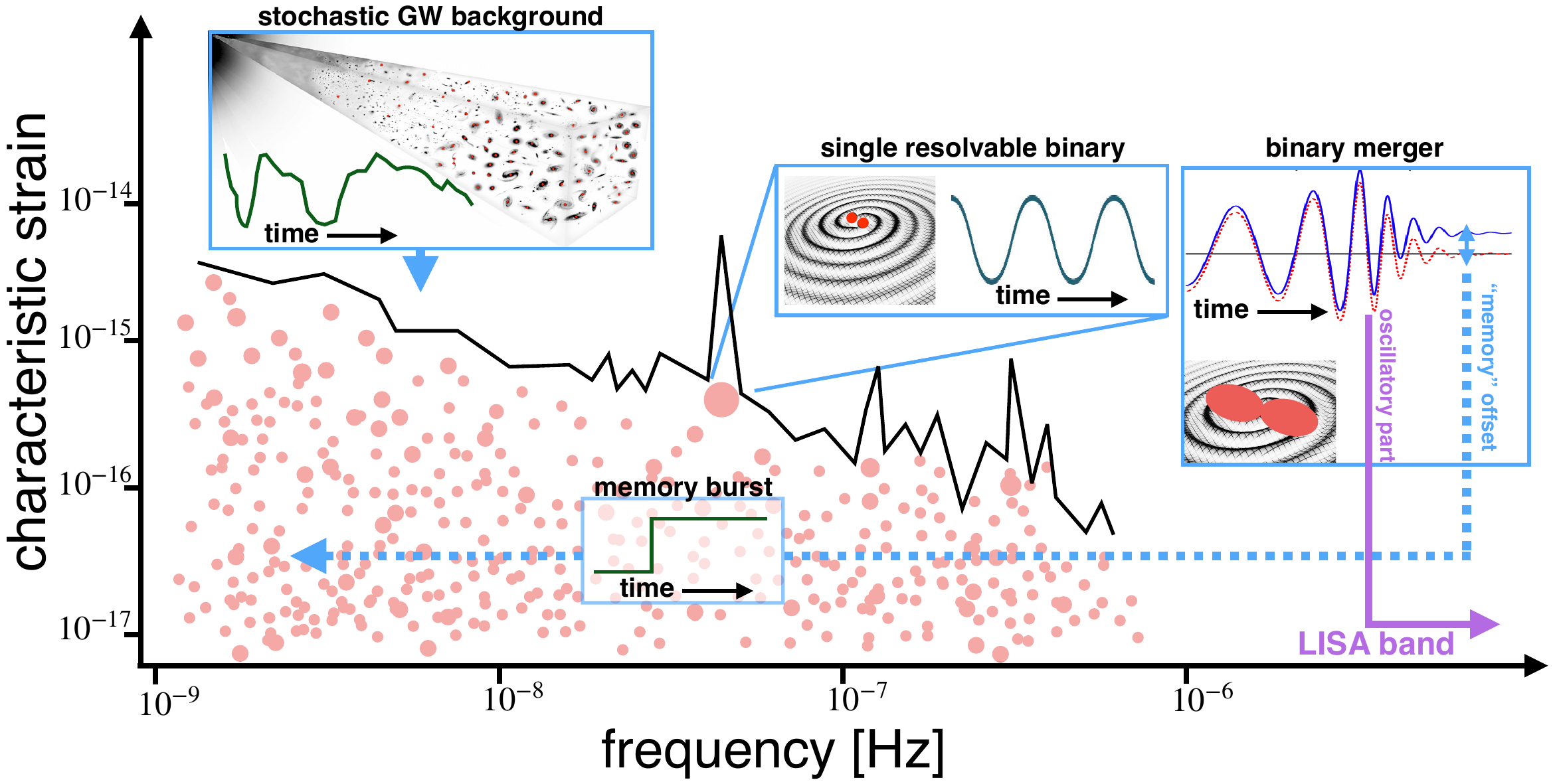}
\vspace{-4mm}
        \caption{A population of supermassive binary black holes will influence pulsar-timing arrays through various inter-pulsar correlated signals, e.g. $(i)$ a GW background causing long-timescale correlations; $(ii)$ individual resolvable binary signals; $(iii)$ the non-oscillatory component of a GW memory burst, causing ramps in timing residuals.}
        \label{GW_signals}
\vspace{-3mm}
    \end{figure*}

\vspace{-2mm}
Fig.~\ref{GW_signals} illustrates the main classes of GW signal that are detectable with PTAs.
\vspace{-5mm}

\subsubsection{Stochastic background}
\vspace{-1mm}
The primary target for PTAs is the GW stochastic background formed by the superposition of GWs from an ensemble of SMBH binaries. Assuming a continuous population of circular binaries whose orbital evolution is dominated by the emission of GWs, the background spectrum follows a $h_{\rm c} \propto f^{-2/3}$ power-law. 
Realistic GW spectra resulting from a finite source population show `spikes' caused by bright binaries that dominate over the ensemble signal in a given frequency bin. Simulated spectra also become steeper at $\mu$Hz frequencies, as fewer systems reside there.

Previous sections indicated that binary environments influence GW emission with the potential to stall or hasten orbital evolution. Effects from dynamical friction, stellar hardening and circumbinary disk interactions may dampen the background signal at lower frequencies, leading to a `turnover` below 10 nHz~\cite{2015PhRvD..91h4055S}. Binaries with eccentricity radiate GW energy over a range of orbital-frequency harmonics, spreading the GW energy across many frequency bins and can significantly impact the characteristic GW spectral shape. These influences on the spectral shape are what allow PTAs to constrain the dynamical evolution mechanisms of SMBH binaries, addressing \textbf{Q2}. 
	
A background of GWs induces low-frequency temporal correlations in pulsar-timing datasets, and (most importantly for detection purposes) quadrupolar inter-pulsar correlated deviations to pulse arrival times described by the \textit{Hellings \& Downs curve} \cite{h+d1983} for an Einsteinian GW background. 
 Recent upper limits on the stochastic background have been used to constrain sub-pc SMBH binary candidates from electromagnetic campaigns and systematic surveys \citep{sesana_testing_2018,holgado_pulsar_2018}. Such efforts have demonstrated that even without a direct detection of the background, astrophysically relevant constraints are already being placed on the demographics of the cosmic SMBH binary population. As a result of non-detection, constraints have also been placed on beyond-General-Relativity theories of gravity \citep{cornish+2018} and population anisotropy \citep{mls+2017,taylor+2015}.  
    
\vspace{-5mm}
\subsubsection{Resolvable binary signals}
\vspace{-1mm}
    PTAs are also sensitive to GWs from discrete binary systems in the local Universe. PTAs have placed blind limits on the GW strain from systems in the $<200\,$Mpc Universe (most recently in \cite{bps+2016,agg+18}), and put direct limits on binary candidates (ex.~\cite{jllw2004, schma+16}). Simulations predict that PTAs will resolve individual binaries within the next $5-20$ years \cite{rosado_2015,kelley_2018a,mls+2017}.

    The observed GW signal depends on the source's sky location, distance, chirp mass, inclination angle, and polarization angle (and potentially orbital eccentricity \citep{thgm16}). The distance and chirp mass are degenerate, but this can be broken if a GW's impact on the Earth \emph{and} pulsars can be extracted (the so-called ``pulsar term'' and ``Earth term''). The GW frequency in these terms will differ due to the evolution of the binary, which depends on the chirp mass but not the system's distance.
\vspace{-5mm}
\subsubsection{Bursts with Memory}
\vspace{-1mm}
    Memory is a non-oscillatory component of GW signals that grows throughout the entire history of a source \citep{christodoulou1991,thorne1992}. During SMBH binary coalescence, when GW emission is maximal, the memory grows quickly, acting as a propagating DC shift in the ambient space-time \citep{favata2009}. Such a GW burst with memory (BWM) is potentially detectable by PTAs \citep{s2009, vHl2010, pbp2010, cj2012, mcc2014}.
    
    When a BWM wave front passes a pulsar, the sudden change in the gravitational potential causes an apparent change in the rotational frequency of the pulsar.  When a BWM wave front passes the Earth, the observed rotation of all pulsars in a PTA will simultaneously change.  The observed change will vary from pulsar to pulsar depending on the relative sky position of the source, following the characteristic quadrupolar pattern of GWs.

\vspace{-5mm}
\section{Key Detectors \& Requirements}\label{sec:detectors}
\vspace{-5mm}
\subsection{Gravitational-wave detection}
\vspace{-1mm}
A detection of the quadrupolar signature of gravitational waves is \emph{the most direct way} to provide evidence for a SMBHB. PTAs will detect SMBHBs in the latest phase of their evolution, at typical separations (depending on mass) $0.001<a<0.1$\,pc, with sensitivity most prevalent to systems of mass $>10^8\,\msun$. If any source evolution (``chirping'') can be detected through either a sufficiently long observation or via the pulsar-term signals, PTAs will be able to \emph{directly track the inspiral evolution of a binary SMBH.}

Detection of the GW background due to SMBHBs is likely going to occur early next decade (if not before). While the initial detection will constrain the amplitude of the background, within $5-10$ years post-detection we expect PTAs to begin to reveal the shape of the background spectrum. Both the amplitude and shape of the background will provide critical constraints for the \emph{en masse} interactions and co-evolution of galaxies with their environments \citep{sesana+2009,ss16,taylor+2017}. To ensure the continued success of these endeavours, PTA collaborations like NANOGrav need access to big-dish radio telescopes (like Arecibo and the Green Bank Telescope) or dish-arrays with equivalent sensitivities (such as DSA2000 or ngVLA) with which to monitor $\gtrsim 50$ pulsars every few weeks over a timescale of years to decades. So equipped, PTAs will stake out the next GW frontier in the 2020s, years before $3^\mathrm{rd}$-generation ground-based GW detectors or the LISA mission become a reality.

\vspace{-5mm}
\subsection{Electromagnetic detection}
\vspace{-3mm}
Ideally, one would be able to directly constrain late-merger binary SMBH orbital dynamics through either the direct tracking of a binary, or through statistical inference based on a sample population. However, both of these are not yet possible due to the paucity of confirmed binary SMBH candidates. Many proposed binary signatures exist (double-peaked emission lines and periodic variability being among the chief suggestions), and have in fact been identified in large galaxy monitoring samples \cite{eracleous+12,Graham2015}. These targets hold some promise, but there are potential alternate physical origins for these types of emission. Thus, most binary SMBH signatures require confirmation. Upcoming synoptic instruments like LSST will perform surveys that will potentially identify hundreds to thousands of such candidates; however, decades-long orbital periods require extended monitoring over at least half a period.

Several upcoming instruments and facilities will contribute critical capabilities to this science, addressing \textbf{Q3}. Among these:
LSST may identify many SMBHB candidates based on periodic variability; wide-field space-based X-ray missions may likewise identify periodic sources; the ngVLA (if fitted with $\sim$8000\,km baselines) and other ongoing long-baseline radio facilities like the VLBA may identify and track the orbits of dual SMBHB cores \citep{rodriguez+06,bansal+17}.

\vspace{-7mm}
\section{Summary}
\vspace{-4mm}
Precision timing of large arrays ($\gtrsim 50$) of millisecond pulsars will detect the nanohertz GW emission from a population of supermassive binary black holes within the next $\sim 3-7$ years. Resolvable individual binary signals and non-oscillatory merger-memory signals are expected to follow $\sim 5$ years thereafter. Long-term monitoring of Galactic millisecond pulsars (as currently undertaken by NANOGrav, the European PTA, the Parkes PTA, and the fused efforts in the form of the International PTA) requires big-dish radio instruments like the Arecibo and Green Bank Telescopes, and dish-arrays with equivalent sensitivities. When combined with pan-chromatic electromagnetic signatures (e.g.\ quasar variability) measured by large synoptic time-domain surveys, VLBI radio imaging, and space-based X-ray missions, the demographics and dynamics of supermassive binary black holes will be unveiled to an unprecedented level. 

\bibliographystyle{unsrt}
\bibliography{bib.bib}

\end{document}